\documentclass[12pt]{article}
\pdfoutput=1
\usepackage{amsmath, pdfsync, hyperref}
\usepackage{amsfonts}
\usepackage{amssymb}
\usepackage{color}
\usepackage{graphicx}
\usepackage{float}
\usepackage[font=footnotesize,labelfont=bf]{caption}

\input xy
\xyoption{all}
\xyoption{web}

\newlength{\abstractwidth}

\renewcommand{\thefootnote}{\fnsymbol{footnote}}
\renewcommand{\thanks}[1]{\footnote{#1}}
\newcommand{\starttext}{
\setcounter{footnote}{0}
\renewcommand{\thefootnote}{\arabic{footnote}}}

\newcommand{\bea}{\begin{eqnarray}}
\newcommand{\eea}{\end{eqnarray}}
\newcommand{\ee}{\end{equation}}
\newcommand{\be}{\begin{equation}}

\newcommand{\reals}{\mathbb{R}}

\DeclareMathOperator{\tr}{tr}

\begin{document}
\starttext
\setcounter{footnote}{0}

\bigskip

\begin{center}

{\Large \bf Holographic RG-flows and  Boundary CFTs}
\vskip 0.6in

{ \bf  Michael Gutperle$^a$  and Joshua Samani$^a$}

\vskip 0.2in

 ${}^a${\sl Department of Physics and Astronomy }\\
{\sl University of California, Los Angeles, CA 90095, USA}\\
{\tt \small gutperle@physics.ucla.edu, jsamani@physics.ucla.edu }

\end{center}

\vskip 0.2in

\begin{abstract}
  Solutions of $(d+1)$-dimensional gravity coupled to a scalar field are obtained which holographically realize interface and boundary CFTs. The solution utilizes a Janus-like $\mathrm{AdS}_d$ slicing ansatz and corresponds to a deformation of the CFT by a spatially-dependent coupling of a relevant operator.  The BCFT solutions are singular in the bulk, but physical quantities such as the holographic entanglement entropy can be calculated.
\end{abstract}

%\newpage
%
%\tableofcontents
%
%\newpage

\baselineskip=17pt
\setcounter{equation}{0}
\setcounter{footnote}{0}

\newpage

%%%%%%%%%%%%%%%%%%%%%%%%%%%%%%%%%%%%%%%%%%%
%%%%%%%%%%%%%%%%%%%%%%%%%%%%%%%%%%%%%%%%%%%
\section{Introduction}
\setcounter{equation}{0}
\label{sec1}
%%%%%%%%%%%%%%%%%%%%%%%%%%%%%%%%%%%%%%%%%%%
%%%%%%%%%%%%%%%%%%%%%%%%%%%%%%%%%%%%%%%%%%%

The AdS/CFT correspondence relates string theory or M-theory on $\mathrm{AdS}_{d+1} \times M $ to a conformal field theory in $d$-dimensions
\cite{Maldacena:1997re,Gubser:1998bc,Witten:1998qj}. The best-known example is given by the duality between Type IIB string theory on $\mathrm{AdS}_5\times S^5$
and 4-dimensional ${\cal N}=4$ super Yang-Mills theory.

It is possible to deform $d$-dimensional conformal field theories by the introduction of boundaries or defects/interfaces such that a subgroup of their (super)conformal symmetries is unbroken. The classification and construction of such interface and boundary
CFTs is an important problem  that enjoys several physical applications
(see e.g. \cite{Cardy:1989ir} for a discussion of the two-dimensional case).

In the context of the AdS/CFT correspondence, it is often possible to find a holographic solution corresponding to the deformation of the CFT.  A well-known example consists of putting a black hole in the bulk of the AdS space which corresponds to a CFT at finite temperature \cite{Witten:1998zw}.

There have been several constructions in the literature of holographic duals to interface CFTs.  In the probe approximation, holographic defects can be described by placing branes  with an $\mathrm{AdS}_d$ world-volume inside the bulk of $\mathrm{AdS}_{d+1}$ \cite{Karch:2000gx}.
The  Janus solution utilizes an ansatz where $\mathrm{AdS}_{d+1}$ is sliced using $\mathrm{AdS}_d$ factors.
The solution  found in  \cite{Bak:2003jk} (see also \cite{Clark:2004sb,Papadimitriou:2004rz}) is dual to an interface of  ${\cal N}=4$  super Yang-Mills theory where the value of the Yang-Mills coupling jumps across the interface. The original solution breaks all supersymmetries, but many generalizations have been found which realize superconformal interface theories \cite{D'Hoker:2007xy,D'Hoker:2007xz,D'Hoker:2008wc,D'Hoker:2008qm,D'Hoker:2009my,Chiodaroli:2011nr,Chiodaroli:2009yw}. For
related  work  by other authors  see \cite{Clark:2005te,Gomis:2006cu,Lunin:2006xr,Lunin:2007ab,Yamaguchi:2006te,Gomis:2006sb,Kumar:2002wc,Kumar:2003xi,Lunin:2008tf}.

Recently, a proposal for the holographic description of boundary CFT has been made in \cite{Takayanagi:2011zk,Fujita:2011fp} (building on the original proposal of \cite{Karch:2000gx}), where an additional boundary $Q$ in the bulk of $\mathrm{AdS}_{d+1}$ cuts off the the bulk spacetime. The intersection of $Q$ with the  boundary of  $\mathrm{AdS}_{d+1}$ constitutes the location of the boundary of the CFT.
%\textcolor{red}{and coincides with the boundary of the asymptotic AdS boundary}.
% In its simplest formulation the construction utilizes the $\mathrm{AdS}_d$ slicing of $\mathrm{AdS}_{d+1}$.
%
% %\textcolor{red}{The space is cut off at a finite value of the slicing coordinate where a brane is located and where one imposes boundary conditions on the bulk fields; this can be thought of as holographically  realizing the degrees of freedom on the boundary of the CFT.}

The half-BPS interface solutions can also be used to obtain holographic duals of BCFTs by taking limits of the regular interface solutions. These solutions were constructed  for four dimensional BCFTs \cite{Assel:2011xz,Aharony:2011yc,Benichou:2011aa}
 and   for two dimensional BCFTs \cite{Chiodaroli:2011fn}. Note that these solutions are necessarily singular.  Recently a completely regular holographic BCFT in two dimensions was constructed using the higher genus half-BPS interface solutions of six dimensional supergravity \cite{Chiodaroli:2012vc}.

Renormalization group flows of CFTs are obtained by deforming the theory by a relevant operator in the UV. The endpoint of the flow in the IR can be another CFT or a massive theory.  In the context of  AdS/CFT, a simple realization of renormalization group flows  is given by turning on a scalar field dual to a relevant operator deformation and solving the coupled equations of motion in the bulk.  Examples of flows between two conformal fixed points  and flows to massive theories can be found in \cite{Freedman:1999gp,DeWolfe:1999cp,Girardello:1999bd}.
Note that on the gravity side, the RG solutions corresponding to the flow to massive theories are generically singular.

The goal of the present paper is to use the techniques of holographic RG flows with the Janus ansatz to find new realizations of boundary and interface CFTs.  We turn on scalar field dual to a relevant operator using a Janus-like $\mathrm{AdS}_d$ slicing of $\mathrm{AdS}_{d+1}$.  An  new feature of our paper is that on the CFT side such an ansatz corresponds to turning on a relevant operator with a source that depends on the coordinate transverse to the interface/boundary.

We obtain numerical solutions of the $(d+1)$-dimensional equations of motion which  realize interfaces, and we find that the solutions interpolate between different values of the source and expectation value of the operator on either side of the interface. Furthermore, we realize boundary CFTs where the solution becomes singular in the bulk. We interpret this as a flow where the source of the operator becomes infinite and the theory on one side of the interface becomes massive leaving only a boundary CFT on the other side of the interface. We illustrate this with example in $d=2$ and $d=4$ dimensions.

The organization of the paper is as follows: In section \ref{sec2} we set up the equations of motion for a scalar coupled to gravity in $d+1$ dimensions  for a Janus ansatz, and we discuss the boundary conditions which correspond to a spatially dependent source for a relevant operator. In section \ref{sec3} the equations of motion are solved numerically, and examples of interface CFTs as well as boundary CFTs are presented.  In section \ref{sec4} we evaluate the entanglement entropy following the prescription of \cite{Ryu:2006bv,Ryu:2006ef}
 for the solutions found in section \ref{sec3}. We discuss our results and possible directions for further research in section \ref{sec5}.

%%%%%%%%%%%%%%%%%%%%%%%%%%%%%%%%%%%%%%%%%%%
%%%%%%%%%%%%%%%%%%%%%%%%%%%%%%%%%%%%%%%%%%%
\section{AdS slicing and BCFT}
\setcounter{equation}{0}
\label{sec2}
%%%%%%%%%%%%%%%%%%%%%%%%%%%%%%%%%%%%%%%%%%%
%%%%%%%%%%%%%%%%%%%%%%%%%%%%%%%%%%%%%%%%%%%

The action for a scalar minimally coupled to $d+1$ dimensional gravity on a manifold $M$ is
\begin{align}
    S &= \int_M d^{d+1}x\sqrt{|g|}\Big(-\frac{1}{4}R +\frac{1}{2}g^{\mu\nu}\partial_\mu\phi\partial_\nu\phi+V(\phi)\Big)
\end{align}
where we have set Newton's constant equal to one. The stress tensor takes the following form
\begin{align}
    T_{\mu\nu}
    &=\partial_\mu\phi\partial_\nu\phi -\frac{1}{2}g_{\mu\nu}g^{\rho\sigma}\partial_\rho\phi\partial_\sigma\phi - g_{\mu\nu} V(\phi).
\end{align}
The equations of motion for the coupled scalar-gravity system are
\begin{align}
    0&= \Delta\phi - V'(\phi) \label{scaleq}  \\
    0&= R_{\mu\nu}-{1\over 2} g_{\mu\nu}R - 2T_{\mu\nu}. \label{graveq}
\end{align}
We normalize the potential by extracting the cosmological constant
\be
V(\phi)= -{d(d-1)\over 4} +\widehat V(\phi).
\ee
We take $\widehat V(0)=0$, so for $\phi=0$ the equations of motion are solved by $\mathrm{AdS}_{d+1}$ with unit curvature radius, where the metric in Poincar\'{e} coordinates is given by
\be\label{adspoinc}
ds^2= {1\over z^2} \Big( dz^2+dx_{\perp}^2-dt^2 + \sum_{i=2}^{d-1} dx_i^2\Big).
\ee
In contrast, The Janus ansatz uses a deformation of the $\mathrm{AdS}_{d}$ slicing of $\mathrm{AdS}_{d+1}$. The Poincar\'{e} slicing \eqref{adspoinc} can be mapped to the $\mathrm{AdS}_d$ slicing by\footnote{The AdS slicing is related to the one of \cite{Bak:2003jk} by a shift in $\mu$ by $\pi/2$. The present coordinates are more convenient for the description of BCFT.}
\be\label{adsmap}
x_{\perp} = y \cos \mu, \qquad z= y \sin \mu,
\ee
which  gives the metric
\be\label{janusmet}
ds^2= {1\over \sin^2\mu}\left( d\mu^2  + { dy^2 -dt^2+ \sum_{i=2}^{d-1} dx_i^2\over y^2}\right).
\ee
Here the slicing coordinate $\mu\in [0,\pi]$.  The boundary of the Poincar\'{e} slicing metric (\ref{adspoinc}) is located at $z=0$. In the AdS slicing, the boundary is mapped into three connected components that we conceptually distinguish from one-another, namely $\mu=0,\pi$ corresponding to two $d$-dimensional half spaces and $y=0$ corresponding to a $(d-1)$-dimensional interface where the two half-spaces are glued together.

\subsection{Janus ansatz and symmetries}

In constructing a holographic dual to an ICFT or BCFT, we look for a bulk spacetime whose group of isometries is the conformal group of the ICFT or BCFT.  For dimensions $d>2$ the conformal group of $\reals^{1,d-1}$  is $\mathrm{SO}(2,d)$, hence a $\mathrm{CFT}_d$  is expected to exhibit $\mathrm{SO}(2,d)$ invariance.

In this paper we consider  an interface or boundary which is the $\reals^{1,d-2}$ subspace localized at $x^\perp=0$.
By definition, we demand that the field theories are invariant under only those elements of $\mathrm{SO}(2,d)$ which preserve the boundary. The subgroup of conformal transformations that preserve this boundary is precisely the conformal group $\mathrm{SO}(2,d-1)$ of the boundary.  This, in turn, is precisely the isometry group of $\mathrm{AdS}_{d}$.  Therefore in searching for a candidate holographic dual to $\mathrm{BCFT}_d$, we look for a spacetime whose bulk exhibits invariance under the full isometry group of $\mathrm{AdS}_{d}$.

The BCFT symmetries are realized by a Janus ansatz  which is based on an $\mathrm{AdS}_d$ sliced metric. All other fields   have nontrivial dependence  only on the slicing coordinate $\mu$. The bulk therefore has manifest $\mathrm{SO}(2,d-1)$ symmetry as desired.
\be
ds^2= f(\mu)\left( d\mu^2  + { dy^2 -dt^2+ \sum_{i=2}^{d-1} dx_i^2\over y^2}\right), \quad \quad  \phi=\phi(\mu).
\ee
With this ansatz, the scalar equation (\ref{scaleq}) becomes
\be\label{scaleqb}
0=\phi''- f \widehat V'(\phi) +{d-1\over 2} {f'\over f}\phi',
\ee
and the gravitational equations become
\begin{align}
    0&={f''\over f}-{3\over 2} {f' f'\over f^2}+{4\over d-1} \phi'\phi'-2 \label{graveq2}\\
    0&={1\over 4} \phi'\phi' -{d(d-1)\over 32}{f'f'\over f^2}-{d(d-1)\over 8}+{d(d-1)\over 8} f-{1\over 2} f \widehat V. \label{constraint}
\end{align}
Note that equation \eqref{constraint} contains only first order derivatives and can be viewed as  a constraint of the evolution with respect to  the  coordinate $\mu$.

\subsubsection{Perturbative solution}

Let us assume that the potential $\widehat V$ has the form
\begin{align}
    \widehat V(\phi)
    &= \frac{1}{2}m^2\phi^2 + \sum_{k=4}^\infty \frac{\widehat V^{(k)}(0)}{k!}\phi^k. \label{potans}
\end{align}
We find a perturbative solution to the scalar-gravity equations.  Consider an ansatz for $f$ and $\phi$ that takes the form of a formal power series in a parameter $\varepsilon$;
\begin{align}
    f(\varepsilon, \mu) &= \sum_{k=0}^\infty f_{2k}(\mu)\varepsilon^{2k}, \qquad
    \phi(\varepsilon, \mu) = \sum_{k=0}^\infty \phi_{2k+1}(\mu)\varepsilon^{2k+1}. \label{pertexp}
\end{align}
Notice that to zeroth order in $\varepsilon$ the solution corresponds to the vacuum solution with no scalar present.  The first order solution corresponds to a small scalar living in the vacuum, the second order solution gives backreaction of the scalar on the gravity solution, and so on.  Plugging this ansatz into the scalar and constraint equations, and setting the coefficients of $\varepsilon$ to zero order-by-order, we obtain a sequence of equations that can, in principle, be solved recursively for the coefficient functions $f_{2k}$ and $\phi_{2k+1}$ in the expansions \eqref{pertexp}.  To zeroth order in $\varepsilon$, we obtain the following equation:
\begin{align}
    0&=(f_0')^2+4 f_0^2-4f_0^3. \label{fzeroeq}
\end{align}
The solution to this equation gives the AdS vacuum. The higher order functions $f_2,f_4, f_6$ and so on give modifications to $f$ due to back-reaction.  The solution to \eqref{fzeroeq} with initial condition $f_0(\pi/2)=1$ is
\begin{align}\label{fzero}
    f_0(\mu) = \frac{1}{\sin^2\mu}
\end{align}
which, as expected, is precisely the appropriate $f$ for the $\mathrm{AdS}_d$ slicing of $\mathrm{AdS}_{d+1}$; see \eqref{janusmet}. To first order in $\varepsilon$, one obtains the following equation:
\begin{align}
    \qquad 0&= 2 f_0 \phi_1''+ (d-1) f_0' \phi_1'-2f_0^2 V''(0) \phi_1.
\end{align}
Plugging in $f_0=1/\sin^2\mu$ and $V''(0)=m^2$ (see \eqref{potans}), we obtain
\begin{align}\label{linearieq}
    0&= \phi_1''-(d-1) \cot \mu \,\phi _1'-m^2  \csc ^2\mu \,\phi _1
\end{align}
whose general solution is a linear combination of the following form
\begin{align} \label{twoline}
  \phi(\mu)&=  C_1~P_{\frac{d-2}{2}}^{\frac{1}{2} \sqrt{d^2+4 m^2}}(\cos \mu ) \sin ^{d/2} +  C_2~  Q_{\frac{d-2}{2}}^{\frac{1}{2} \sqrt{d^2+4 m^2}}(\cos \mu ) \sin ^{d/2}\mu.
\end{align}
In the following we will not go to higher than first order in $\varepsilon$ since we will solve the equations numerically.

\subsection{Holographic dictionary}
The standard holographic dictionary relates
the mass $m$ of the scalar field to the conformal dimension $\Delta$  of the dual operator $O_{\Delta}$;
\begin{equation}
    m^2 = \Delta(\Delta-d), \quad \quad \Delta={1\over 2}\Big( d+ \sqrt{d^2+4m^2}\Big).
\end{equation}
The second relation holds for the so-called ``standard quantization."  We will consider operators which are IR relevant.  This is equivalent to considering scalar fields with squared mass $m^2$ satisfying
\begin{equation}\label{rangemass}
-{d^2\over 4} <m^2<0, \qquad \frac{d}{2} < \Delta <d.
\end{equation}
Near the AdS boundary in Poincar\'{e} slicing the the scalar field behaves as follows
\be
\phi(x) \sim  \phi_1(x) z^ {d-\Delta} + \phi_2 (x) z^ \Delta +\cdots.
\ee
The standard holographic dictionary identifies $\phi_1$ with  the (linearized) source added to the action and $\phi_2$ with the expectation value for the operator $O_\Delta$ \cite{Witten:1998qj}.

The  solution of  the linearized scalar equation in the $\mathrm{AdS}$ slicing  (\ref{twoline}) behaves as follows
near the boundary $\mu=0$:
\begin{equation}\label{sourceexp}
   \phi(\mu)\sim \alpha\;  \mu^\Delta + \beta\; \mu^{d-\Delta}+\cdots.
\end{equation}
The constants $\alpha,
\beta$ determine the initial conditions for  the evolution equations \eqref{scaleqb} and \eqref{graveq2}. One might conclude, by following the  holographic prescription  outlined above  that $\beta $ corresponds to a constant source  and $\alpha $ to a constant expectation value  of the dual operator on the half space located at $\mu=0$.    However, inverting the relations (\ref{adsmap}) for $\mu \to 0$ and $y>0$ we obtain
\be
\mu =  {z\over x_{\perp}}, \quad  y=x_{\perp}
\ee
which is valid for small $\mu$. Hence by mapping coordinates from AdS slicing   to Poincar\'{e} slicing, one obtains the behavior near $z=0$ but with $x_\perp>0$ which corresponds to the points away from the interface;
\begin{equation}\label{mapjanus}
\lim_{z\to 0}   \phi(z,x_{\perp})\sim  {\alpha \over (x_{\perp})^{\Delta} } z^\Delta + {\beta\over( x_{\perp})^{d-\Delta} } z^{d-\Delta}+\cdots.
\end{equation}
In the Poincar\'{e} slicing realization of RG flows, the surfaces of constant $z$  correspond to a fixed energy  scale in the dual CFT.  It follows from \eqref{mapjanus} that the scalar behavior near the boundary  corresponds to sources and expectation values for the dual operator which are dependent on the transverse coordinate $x_{\perp}$.
In a recent papers  \cite{Dong:2012ena,evanew}  spacetime dependent couplings in RG-flows were discussed and it was noted that the space time dependence can change the relevance of the operator perturbation.
 This is a new feature of the Janus ansatz and is not the case for  holographic RG flows with Poincar\'{e} symmetry which are translation invariant along all directions in $\reals^d$. The extra $x_{\perp}$ dependence of the coupling in (\ref{mapjanus}) seems to make the perturbation marginal, we will still call the evolution RG-flow as the coupled scalar-gravity evolution shares many features of the holographic  Poincar\'{e} RG-flow.

%%%%%%%%%%%%%%%%%%%%%%%%%%%%%%%%%%%%%%%%%%%
%%%%%%%%%%%%%%%%%%%%%%%%%%%%%%%%%%%%%%%%%%%
\section{Holographic ICFT and BCFT via RG flow}
\setcounter{equation}{0}
\label{sec3}
%%%%%%%%%%%%%%%%%%%%%%%%%%%%%%%%%%%%%%%%%%%
%%%%%%%%%%%%%%%%%%%%%%%%%%%%%%%%%%%%%%%%%%%

Turning on a relevant operator in a $d$-dimensional CFT   generates a renormalization group flow. In the IR, the theory can flow either to a new conformal fixed point, or become massive. The holographic realization of RG flows  in the Poincar\'{e} slicing has been studied many papers (see e.g.  \cite{Freedman:1999gp,DeWolfe:1999cp}). In the following we will instead study RG flows using the Janus ansatz described in the previous section.

The main difference between the Poincar\'{e} slicing and $\mathrm{AdS}_d$ slicing lies in the fact  that for the  $\mathrm{AdS}_d$ slicing, as discussed in  section \ref{sec2}, the UV boundary of $\mathrm{AdS}_{d+1}$ has three components, corresponding to the two $d$-dimensional half spaces glued together at a $d-1$-dimensional interface.

 For a holographic interface  we choose the two boundary components associated with the two  half spaces to be located at $\mu_+$ and $\mu_-$. Since the metric becomes asymptotically AdS at $\mu=\mu_\pm$,  the metric and scalar field
behave as follows:
\begin{align}
\lim_{\mu \to \mu_\pm} f(\mu) &\sim  {1 \over (\mu- \mu_\pm)^2} + \cdots\notag\\
\lim_{\mu \to \mu_\pm} \phi(\mu) &\sim  \alpha_\pm (\mu-\mu_\pm)^\Delta + \beta_\pm (\mu-\mu_\pm)^{d-\Delta}+\cdots.
\end{align}
It follows from \eqref{mapjanus} that $\alpha_\pm$ correspond to position-dependent expectation values for $O_\Delta$ at on the two half spaces and $\beta_\pm$ correspond to the position-dependent sources for $O_\Delta$ at on the two half spaces.  Note
that for a smooth solution of the equations of motion (\ref{scaleqb})  and (\ref{graveq2}) only two
of the four constants $\alpha_\pm,\beta_\pm$ are independent. In particular for the linearized
solution given in  (\ref{twoline}), one neglects the gravitational back reaction and hence the value of $\mu_\pm$ is unchanged from the undeformed AdS values, i.e. $\mu_- =0$ and $\mu_+ =\pi$. From (\ref{twoline})  one can read off a linear relation between $a_+, \beta_+$ and $a_-, \beta_-$. For simplicity and in order to compare with the boundary conditions that we impose on numerical solutions in later sections, we choose to set the expectation value to zero at $\mu_-$, namely we choose $\alpha_-=0$.  The resulting linear relations between $\alpha_+$, $\beta_+$, and $\beta_-$ are

%\begin{eqnarray}
%\alpha_+ &=& d \alpha_-  + e  \beta_- \nonumber \\
%\beta_+ &=& f \alpha_- + g  \beta_-\label{linearab}
%\end{eqnarray}

%\begin{align}
%    \begin{pmatrix}
%      \alpha_+ \\
%      \beta_+ \\
%    \end{pmatrix}=
%    \begin{pmatrix}
%      d & e \\
%      f & g \\
%    \end{pmatrix}
%    \begin{pmatrix}
%      \alpha_- \\
%      \beta_- \\
%    \end{pmatrix}
%\end{align}

\begin{align}
\alpha_+ &=  \frac{\pi  2^{d-2 \Delta } \csc \left(\frac{1}{2} \pi  (d-2 \Delta )\right) \Gamma \left(\frac{d}{2}-\Delta +1\right)}{\Gamma (1-\Delta ) \Gamma (d-\Delta ) \Gamma \left(-\frac{d}{2}+\Delta +1\right)} \;   \beta_- + \mathcal O ( \beta_- ^2), \label{linearapl} \\
  \beta_+&=  -\sin \left(\frac{\pi  d}{2}\right) \csc \left(\frac{1}{2} \pi  (d-2 \Delta \right)
  \;\beta_- + \mathcal O ( \beta_- ^2).\label{linearbpl}
\end{align}

\begin{figure}[h]
    %Plots generated from notebook "d=2 numerics-6-30" in draft folder.
  \centering
    \includegraphics[scale=0.8]{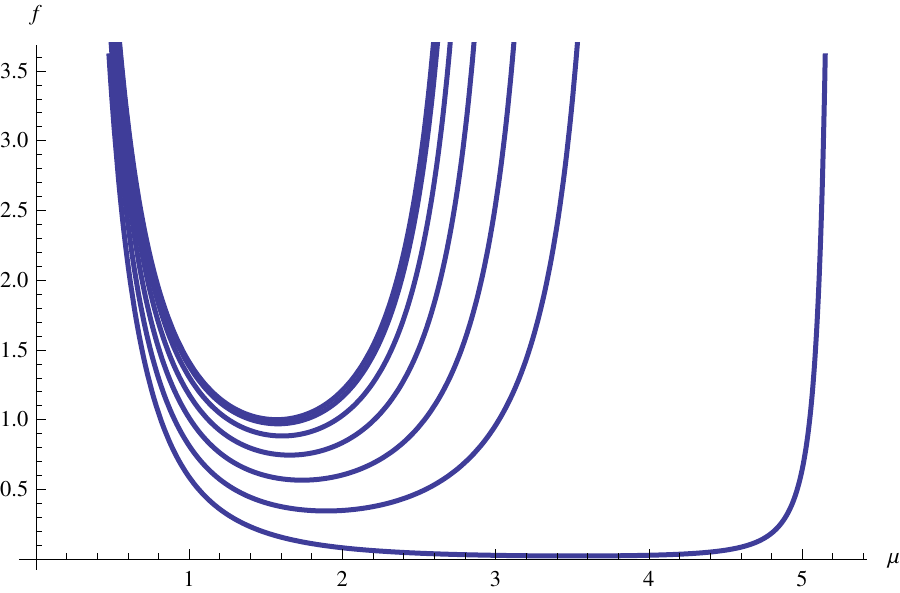} \includegraphics[scale=0.8]{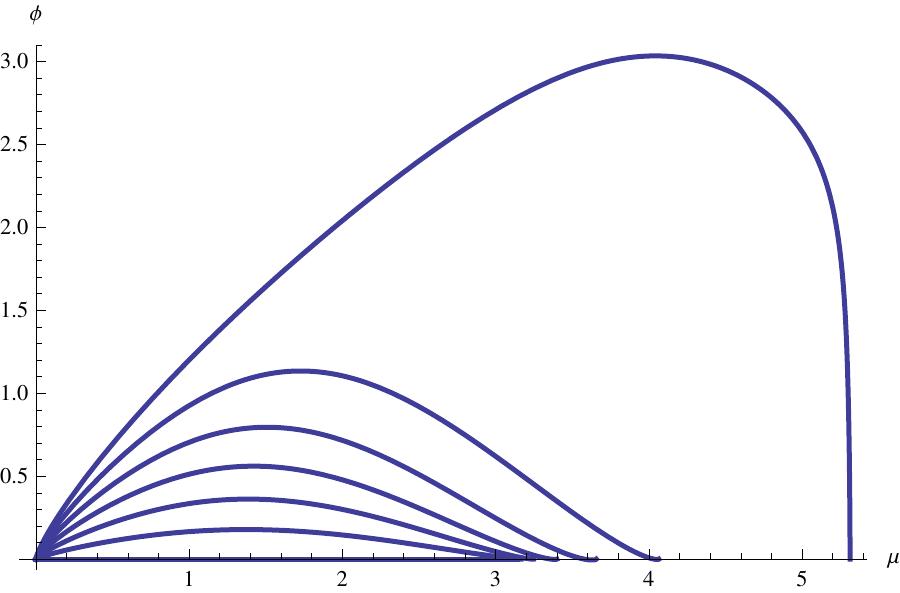} \includegraphics[scale=0.8]{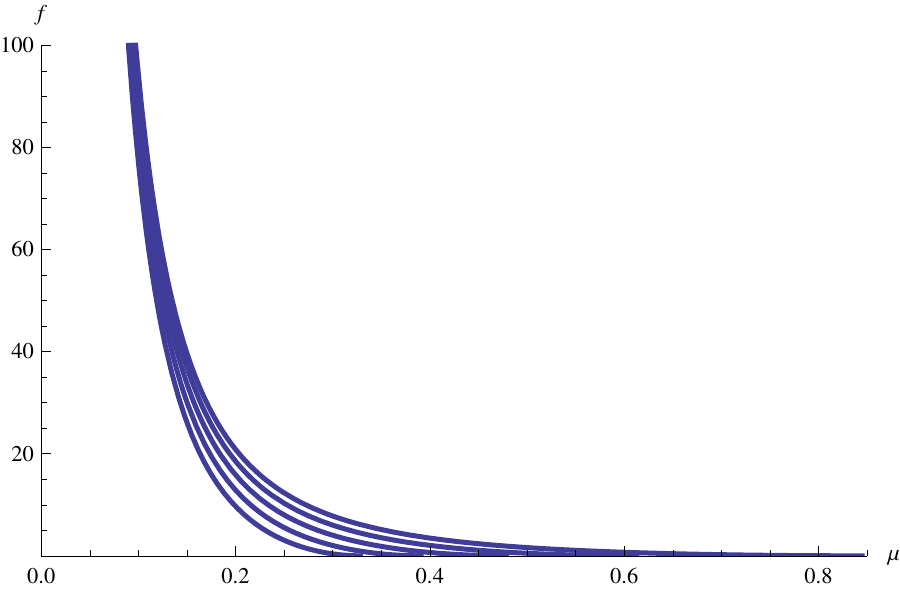} \includegraphics[scale=0.8]{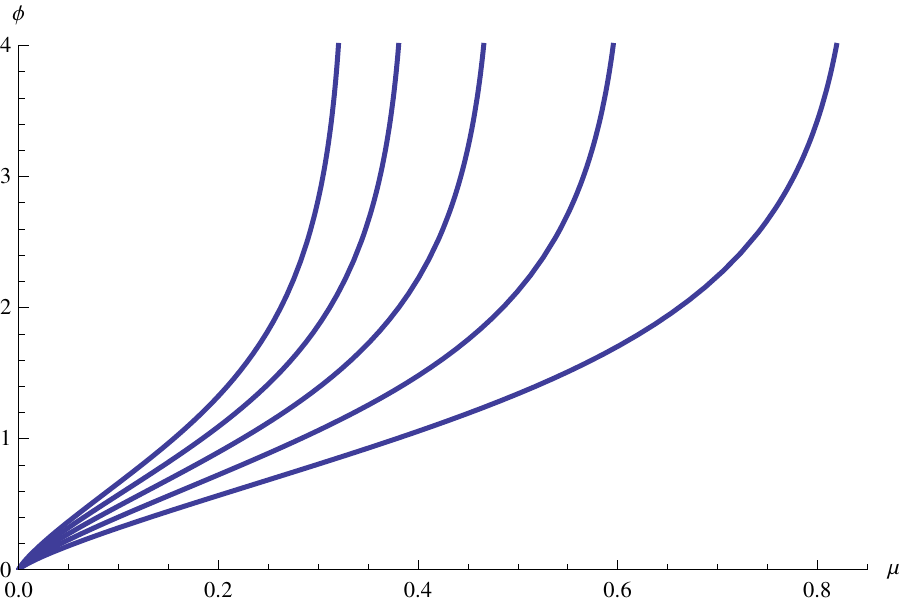}
  \caption{Plots of $f(\mu)$ and $\phi(\mu)$ in the ICFT (top row) and BCFT (bottom row) cases with $\Delta=1.202$ and $\lambda_4=-4.8$.  We have used $d=2$ with potential given by \eqref{hatpot}.  The family of curves in each plot is generated by varying the value of the source $\beta_-$.  The ICFT curves correspond to $\beta_- = 0,0.2,0.4,\dots, 1.2$ while the BCFT curves correspond to $\beta_- = 2,2.5,3,3.5,4$.}
\label{fphiplot2}
\end{figure}

\subsection{ICFT and BCFT in $d=2$}\label{twodnumerics}

In order to go beyond the linearized approximation, we numerically solve the equations of motion \eqref{scaleqb} and \eqref{graveq2} subject to the constraint \eqref{constraint}. We choose to locate one boundary at $\mu_- =0$, and we impose boundary conditions on the scalar field there corresponding to a vanishing expectation value with only a source turned on, i.e. $a_-=0$.   In the following we study the case $d=2$ which corresponds to a deformation of a 2-dimensional CFT. We consider a toy model with a potential
\be
\hat V(\phi) = {1\over 2} \Delta(\Delta-2)\phi^2 + {1\over 4!} \lambda_4 \phi^4. \label{hatpot}
\ee
As an example we choose the the operator $O_\Delta$ to be relevant and have dimension $\Delta=1.2$ and consider a potential with a small negative quartic coupling $\lambda_4=-4.8$.  Note that for these values, $\phi=0$ is the only extremum of the potential $\hat V$. The behavior of the resulting solution depicted in the following is generic for any relevant operator deformation.

As a function of the source $\beta_-$, the numerical solution displays the following properties:
For very small $\beta_-$, the values of $\alpha_+,\beta_+$, which are obtained by a numerical fit, approach their linearized values given by \eqref{linearapl} and \eqref{linearbpl}. The values of the source $\beta_+$ and expectation value $\alpha_+$ on  the second half space  grow for increasing values of the source.  Following the discussion above, we can interpret these solutions as Janus-like interfaces, where the two CFTs defined on the half spaces at $\mu=\mu_\pm$ have different $x_\perp$-dependent sources and expectation values on either side.

At a critical value of $\beta_-$, both $\mu_+$ and $\alpha_+,\beta_+$ diverge, the metric function $f$ approaches a zero, and the solution becomes singular. We interpret this in the following way: The operator deformation on the half-space at $\mu=\mu_+$ becomes so large that the theory is becoming massive, and the second asymptotic region disappears. Consequently for values of the source $\beta_-$ larger than the critical value, the solution becomes singular in the bulk and corresponds to a BCFT since there is only one asymptotically AdS boundary corresponding to a single half-space.

\begin{figure}
\centering
\includegraphics[scale=0.9]{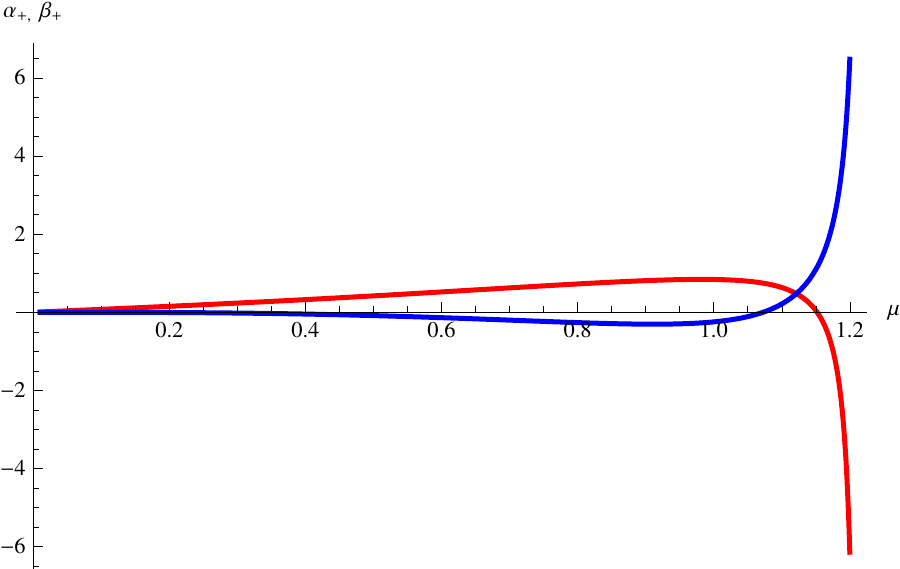}
\caption{ Plot of expectation value  $\alpha_+$ (red)  and source $\beta_+$ (blue) as a function of the source $\beta_-$}
\vspace{-30pt}
\label{abetaplot}
\end{figure}

We will use these numerical solutions to study the entanglement entropy for the ICFT and BCFT solutions in section \ref{sec43}.

\begin{figure}[h]
  \centering
    \includegraphics[scale=0.85]{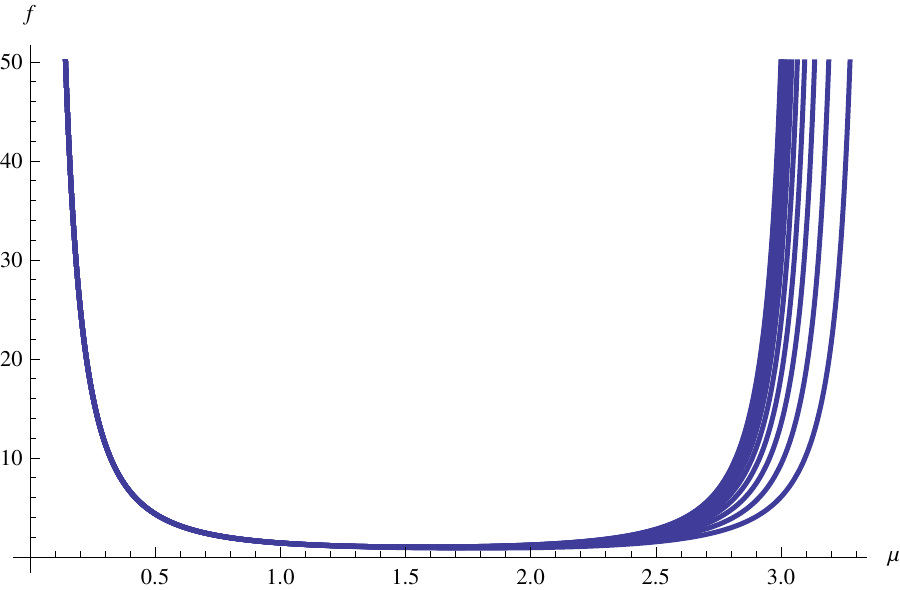} \includegraphics[scale=0.85]{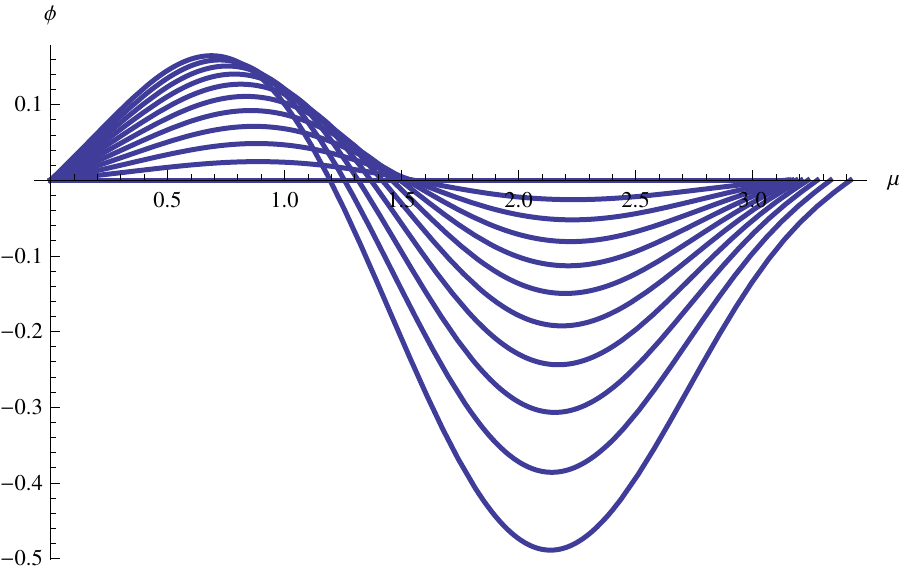}
    \includegraphics[scale=0.85]{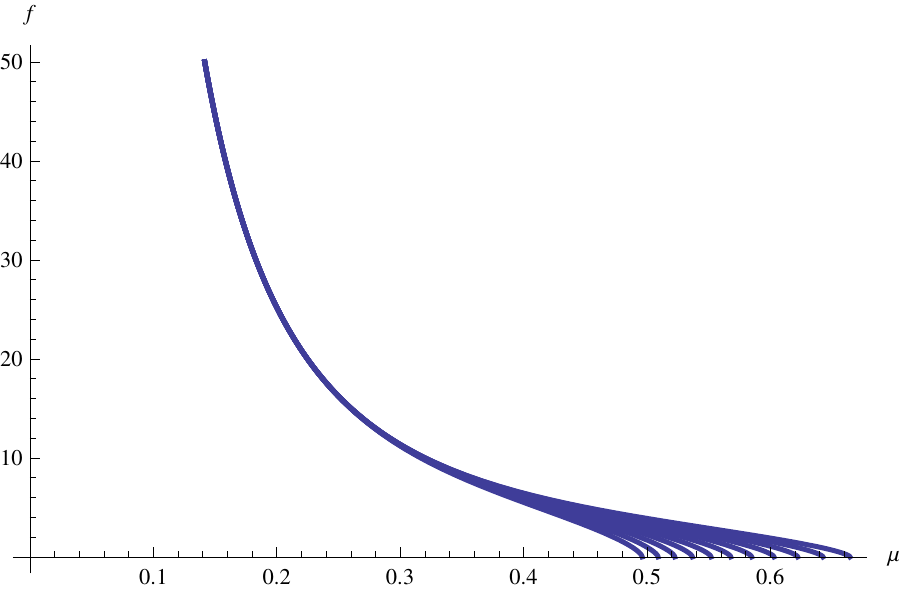} \includegraphics[scale=0.85]{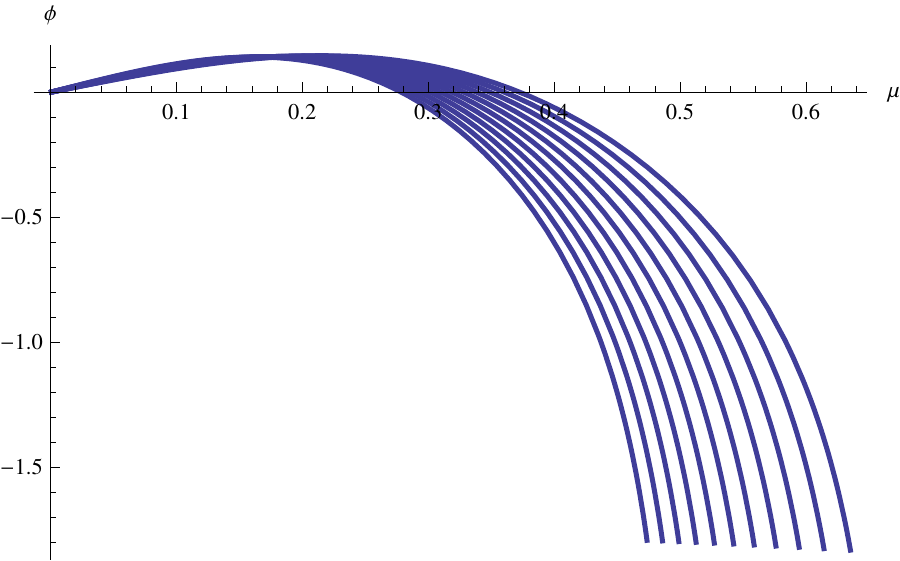}
    %\caption{Plot of scalar $\phi(\mu)$ for GPPZ flow in the BCFT case for varying values of the source $\beta_-$.}
  \caption{Plots of $f(\mu)$ and $\phi(\mu)$ in the ICFT (top row) and BCFT (bottom row) cases.  We have used $d=4$ with GPPZ potential given by \eqref{gppzpot}.  The family of curves in each plot is generated by varying the value of the source $\beta_-$.  The ICFT curves correspond to $\beta_- = 0, 0.03, 0.06, \dots, 0.3$ while the BCFT curves correspond to $\beta_- = 1, 1.03, 1.06, \dots, 1.3$.}
  \label{gppzfphi}
  \vspace{-10pt}
\end{figure}

\subsection{Interface and BCFT in $d=4$}\label{fdourflow}
In this section we consider specific examples of $d=4$ ICFT and BCFT RG-flows. Qualitatively the solutions behave in the same way as in the 2-dimensional case presented in section \ref{twodnumerics}.  We consider a truncation of $\mathcal N=8$ supergravity introduced in \cite{Girardello:1999bd} called the GPPZ solution.  The potential $V$ can be expressed in terms of a pre-potential
\be
  W(\phi) = -{3\over 4} \left[1+ \cosh\left({2\phi\over \sqrt{3}}\right) \right] \label{gppzpot}
\ee
which determines the potential as follows:
\begin{align}\label{dfourpot}
  V(\phi) &= {1\over 2} \left(\partial_\phi W\right)^2-{4\over 3} W^2 \notag \\
  &= -3 -{3\over 2} \phi^2 -{1\over 3} \phi^4 + o(\phi^6).
\end{align}
Expanding the potential around the maximum $\phi=0$ indicates that the scalar field is dual to a relevant  operator with dimension $\Delta=3$. In \cite{Girardello:1999bd} it was argued that a Poincar\'{e} slicing RG flow solution becomes singular and the singularity  represents the flow of $\mathcal N=4$ SYM to a massive fixed point with $\mathcal N=1$ supersymmetry\footnote{See \cite{Liu:2012ee,Albash:2011nq} for a recent discussion of Poincar\'{e} RG flows for the GPPZ flow and the evaluation of entanglement entropy for such flows.}.  In the following, we numerically solve the equations of motion for an AdS-sliced RG-flow in $d=4$ with the  potential  given in eq. (\ref{dfourpot}).

As with the 2-dimensional solutions, we set the expectation value $\alpha_-$ of the dual operator at $\mu_-=0$ to zero, and we plot the corresponding solutions for a few values of the operator source $\beta_-$. Qualitatively, the behavior of the solutions is very similar to that of the solutions found in section \ref{twodnumerics}. For small values of $\beta_-$, we have a holographic ICFT where at the $\mu=\mu_+$ boundary the scalar generally has a nonzero source and expectation value. A set of representative plots is given in figure \ref{gppzfphi}. For a critical value of the source $\beta_-$, the solution becomes singular, and we have a holographic BCFT.  The location of the singularity is a function of $\beta_-$.   A set of representative plots is given in figure \ref{gppzfphi}.

%%%%%%%%%%%%%%%%%%%%%%%%%%%%%%%%%%%%%%%%%%%
%%%%%%%%%%%%%%%%%%%%%%%%%%%%%%%%%%%%%%%%%%%
\section{Entanglement entropy and minimal surfaces}
\setcounter{equation}{0}
\label{sec4}
%%%%%%%%%%%%%%%%%%%%%%%%%%%%%%%%%%%%%%%%%%%
%%%%%%%%%%%%%%%%%%%%%%%%%%%%%%%%%%%%%%%%%%%

\begin{figure}
  \begin{minipage}[c]{0.48\textwidth}
  \centering
    \includegraphics[scale=0.6]{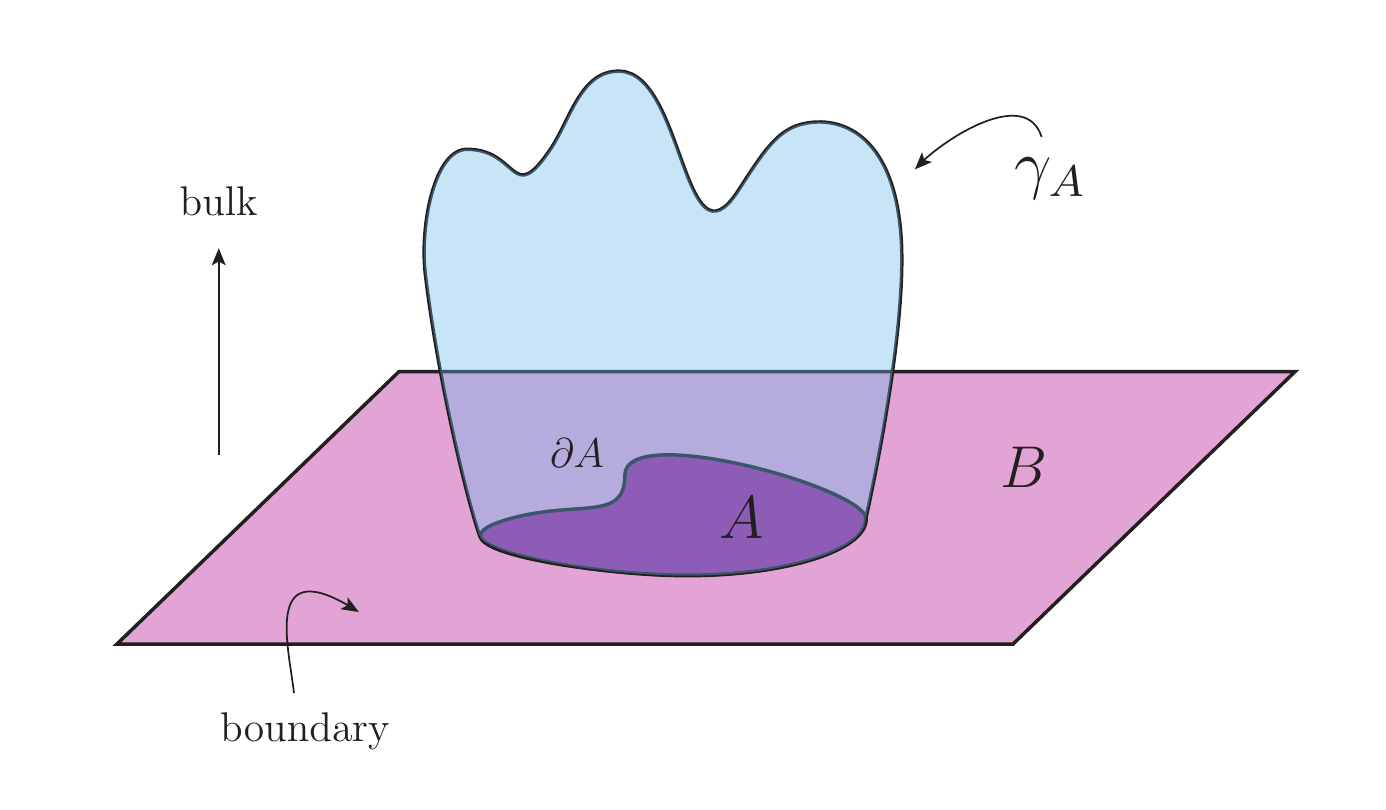}
    \caption{Minimal surface $\gamma_A$ used for the calculation of holographic entanglement entropy.}
    \label{general-entangle}
  \end{minipage}\hfill
  \begin{minipage}[c]{0.48\textwidth}
  \centering
    \includegraphics[scale=0.48]{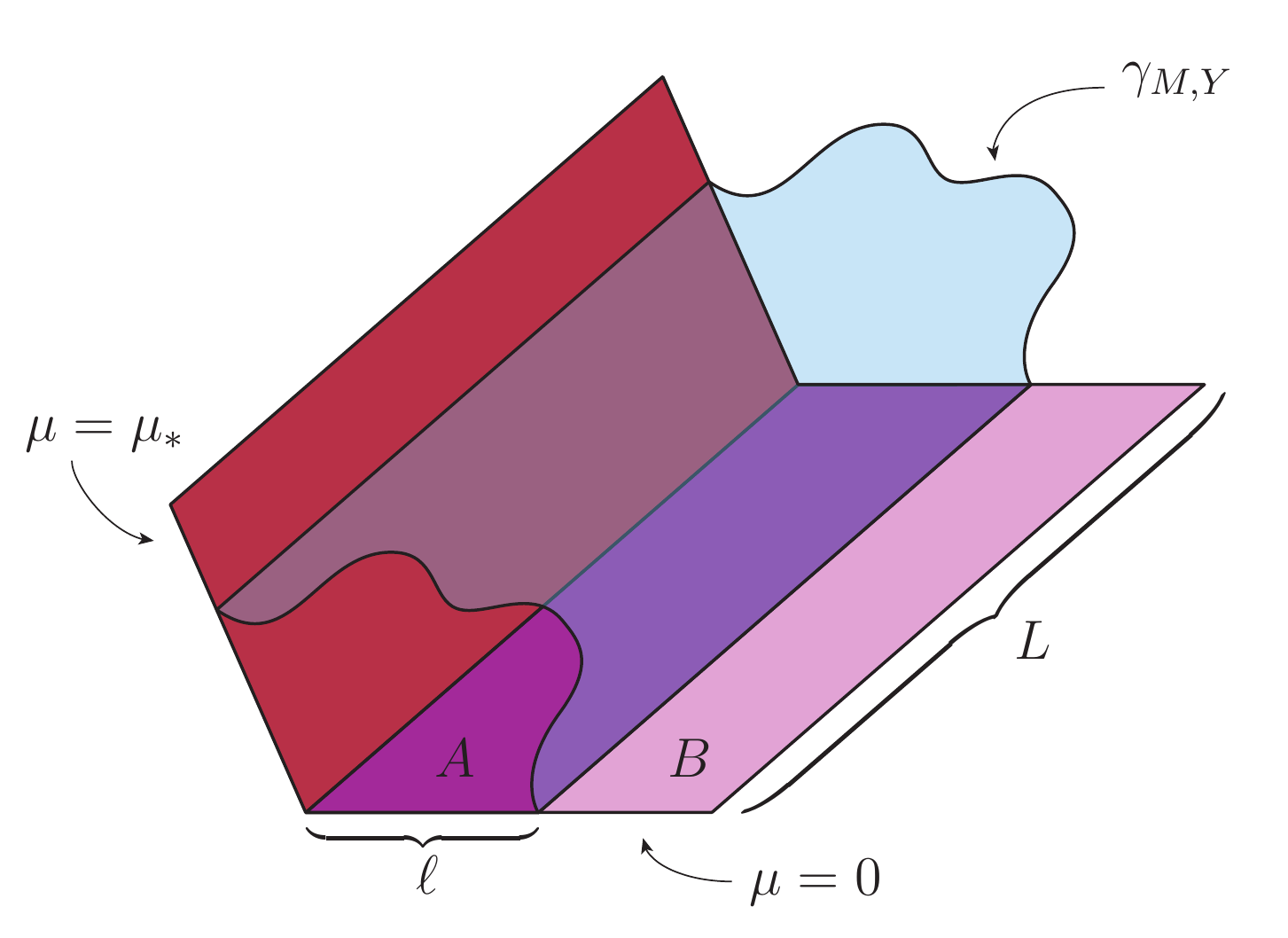}
    \caption{Minimal surface for calculation of the holographic entanglement entropy in the strip geometry in the case of a flow to a BCFT.}
    \label{slab-entangle}
  \end{minipage}
  \vspace{0pt}
\end{figure}

Consider a QFT defined on a $d$-dimensional spacetime, and let $A$ be a subregion of a constant time slice of that spacetime.  The entanglement entropy  (see e.g. \cite{Calabrese:2004eu} for a review) for $A$ is defined as follows.  Let $B$ be the complement of $A$ in the time slice. The Hilbert space of the system can be expressed as a tensor product of degrees of freedom localized in either  $A$ or $B$, namely  $\mathcal{H}=\mathcal{H}_A\otimes\mathcal H_B$.  The general  state of the system can be described by a density operator $\rho$ on $\mathcal H$, and the state of a subsystem  $ A$  is described by a reduced density operator $\rho_A = \tr_B\rho$.  One then defines the entanglement entropy of system $A$ with system $B$ as the von Neumann entropy associated with the reduced density operator $\rho_A$;
\begin{align}
    S_A=-\tr\rho_A\ln\rho_A.
\end{align}
A proposal to holographically calculate the entanglement entropy of $d$-dimensional CFT was discussed in \cite{Ryu:2006bv,Ryu:2006ef}.
Working in Poincar\'{e} coordinates, the CFT is defined on Minkowski space $\reals^{1,d-1}$ which can be
 thought of as the boundary of $\mathrm{AdS}_{d+1}$.
 The subsystem $A$ is a $d$-dimensional sub-region in the constant-time slice.
The boundary of $A$ will be denoted by $\partial A$ (see figure \ref{general-entangle}).
One finds the static minimal surface $\gamma_A$ that extends into the $\mathrm{AdS}_{d+1}$ bulk and ends on $\partial A$
as one approaches the boundary of $\mathrm{AdS}_{d+1}$.
The holographic entanglement entropy can then be calculated as follows  \cite{Ryu:2006bv,Ryu:2006ef}:
\bea
S_A = {{\rm Area}(\gamma_{ A}) \over 4 G^{(d+1)}_{N}}, \label{heemaster}
\eea
where ${\rm Area}(\gamma_A)$ denotes the area of the minimal surface $\gamma_A$, and $ G^{(d+1)}_{N}$ is
the Newton constant of the $(d+1)$-dimensional gravity.

\subsection{Janus minimal surfaces}

We first adapt the holographic entanglement entropy formula \eqref{heemaster} to the Janus geometry in the BCFT case.  Given $\ell>0$, we divide a time slice of the BCFT living on the boundary $\mu=0$ into two regions: region $A$ consisting of all points satisfying $y<\ell$, and region $B$ consisting of all points satisfying $y>0$.  We want to compute the entanglement entropy between these two regions.  Taking a time slice of the Janus metric, we compute the minimal surfaces that intersects $\partial A$ which consists of those points with $\mu=0$ and $y=\ell$.  This setup is called the strip geometry.  A time slice of the Janus metric has the following metric:
\begin{align}
    ds^2 = f(\mu)\left(d\mu^2 + \frac{dy^2 +\sum_{i=1}^{d-2}dx_i^2}{y^2}\right). \label{tsmet}
\end{align}
In the strip geometry, we expect minimal surfaces to be invariant under translations in the transverse directions $\vec x = (x_1, \dots, x_{d-2})$, so we look for minimal surfaces with embedding coordinates of the following form:
\begin{align}
    \mu(s, \vec x) &= M(s), \qquad
    y(s, \vec x) = Y(s), \qquad
    \vec x(s, \vec x) = \vec x.
\end{align}
The induced metric $h_{ij}(s, \vec x)$ on a manifold described by these embedding coordinates is diagonal with entries
\begin{align}
    h_{ss}(s, \vec x)
    &= f(M(s))\left(M'(s)^2+\frac{Y'(s)^2}{Y(s)^2}\right) \\
    h_{ii}(s,\vec x)
    &= \frac{f(M(s))}{Y(s)^2}; \qquad i=1, \dots, d-2.
\end{align}
In the transverse directions, we take the strip to be a cube $[0,L]^{d-2}$ of side length $L$.  The area of a surface $\gamma_{M,Y}$ parameterized in this way is
\begin{align}
    \mathrm{Area}[d,f;\gamma_{M,Y}]
    %&=\int d^{d-2}x\int %ds\,\sqrt{\frac{f(M(s))^{d-1}}{Y(s)^{2d-4}}\left(M'(s)^2+\frac{Y'(s)^2}{Y(s)^2}\right)} \notag\\
    &= L^{d-2} \int ds\, {\cal L}_{d,f;M,Y}(s), \label{aread}
\end{align}
where
\begin{align}
    {\cal L}_{d,f;M,Y}(s) = \sqrt{\frac{f(M(s))^{d-1}}{Y(s)^{2d-4}}\left(M'(s)^2+\frac{Y'(s)^2}{Y(s)^2}\right)} \label{lagd}
\end{align}
can be viewed as the ``Lagrangian" for the area functional.  The problem of finding minimal surfaces is equivalent to determining solutions to the Euler equations for the functions $M$ and $Y$ obtained by minimizing this area functional. The solution will be characterized by a parameterized curve $(M(s), Y(s))$ in the $\mu$-$y$ plane which gives the constant $\vec x$ profile of the surface.  The minimal surface Euler equation obeyed by the component functions $M$ and $Y$ is given by
\begin{align}
  0 &= 2 Y f(M) \left[(d-2) Y^2 \left(M'\right)^3+(d-3) M' \left(Y'\right)^2+Y \left(M' Y''-Y' M''\right)\right] \notag \\
  &\hspace{7cm}+(d-1) Y' f'(M) \left[Y^2 \left(M'\right)^2+\left(Y'\right)^2\right]. \label{msM}
\end{align}

\subsection{Asymptotic Expansion and initial data}

In a sufficiently small neighborhood of the boundary $\mu=0$, any minimal surface intersecting the point $(\mu, y) = (0,\ell)$ can be parameterized as follows $(M(s),Y(s)) = (s, Y(s))$.  In other words, the parameter $s$ is simply the angular coordinate $\mu$.  The minimal surface equation \eqref{msM} then reduces to the following equation for the function $Y$:
\begin{align}
    0
    &=(d-1) f' Y' \left(Y^2+\left(Y'\right)^2\right)+2 f Y \left((d-2) Y^2+(d-3) \left(Y'\right)^2+Y Y''\right). \label{asymms}
\end{align}
In this description, the boundary data are $(0,\ell) = (0, Y(0))$, so in particular, we require the initial datum $Y(0) = \ell$ on the function $Y$.  On the other hand, for the boundary conditions which we are considering, i.e. setting the expectation value of the dual operator to zero, $f$ has the following asymptotic expansion near $\mu=0$
We display the expansion for the two cases we are discussing in the paper. First, for $d=2$ and generic $1<\Delta<2$
\begin{eqnarray}
d=2, \Delta:&&   \phi(\mu) =\beta_- \mu^{2-\Delta} -{1\over 12} \beta_- (\Delta-1) \mu^{4-\Delta} +\cdots  \nonumber  \\
&& f(\mu) = {1\over \mu^2} +{1\over 3} +{1\over 15} \mu^2-2 \beta_-^2{\Delta-2\over  2\Delta-5} \mu^{2-2\Delta},
\end{eqnarray}
and second for  $d=4, \Delta=3$ and the GPPZ potential.
\begin{eqnarray}\label{dfourexp}
d=4,\Delta=3:&&    \phi(\mu) = \beta_- \mu -\beta_- \mu^3 \log(\mu)  +\cdots \nonumber \\
&&f(\mu)= {1\over \mu^2} +{(3-2 \beta_-^2)\over 9} +{2 \beta_-^2\over 5} \mu^2 \log\mu +\cdots.
\end{eqnarray}
In both cases $\beta_-$ is the source of the dual operator and we have set the expectation value to zero. The behavior of the minimal surface function $Y$ near $\mu=0$ can then be obtained  by plugging the expansion of $f$  into \eqref{asymms}, yields the following   expansions
\begin{eqnarray}
d=2, \Delta:&&
    Y(\mu) = \ell  + \hat  y \;  \mu^2  + {\beta_-^2\;  \hat   y  (\Delta-2)\over (\Delta-3)(2\Delta-5)} \mu^{6-2\Delta} +\cdots  \label{msae} \\
    d=4, \Delta=3:&&    Y(\mu) = \ell   + {\ell\over 2} \mu^2 + \hat   y \; \mu^4 + { \beta_-^2 \;  \ell\over 6}   \mu^4\log\mu+\cdots.  \label{msaeb}
\end{eqnarray}
For both cases  the expansion depends on two arbitrary integration constants $\ell,\hat   y$, as is expected for a second order differential equation. The constant $\ell$ determines the location where $Y$ intersects the AdS boundary at $\mu=0$, the second constant $y$ determines (roughly) how the minimal surface curves.

\subsection{Holographic entanglement entropy in $d=2$}\label{sec43}

Setting $d=2$ in the metric \eqref{tsmet} yields
\begin{align}
    ds^2 = f(\mu)\left(d\mu^2 + \frac{dy^2}{y^2}\right). \label{tsmtwo}
\end{align}
%Minimal surfaces used to compute holographic entanglement entropy are geodesics in this geometry. Plugging $d=2$ into the minimal surface equations \eqref{msM} and \eqref{msY} gives the following two equations for such geodesics
%\begin{align}
%    0
%    &=Y' \Big[Y' f'(M) \left(Y^2 \left(M'\right)^2+\left(Y'\right)^2\right)
%    -2 Y f(M) \left(M' \left(Y'\right)^2-Y \left(M' Y''-Y' M''\right)\right)\Big] %\label{2msM}\\
%    0
%    &=M' \Big[Y' f'(M) \left(Y^2 \left(M'\right)^2+\left(Y'\right)^2\right)
%    -2 Y f(M) \left(M' \left(Y'\right)^2-Y \left(M' Y''-Y' M''\right)\right)\Big] %\label{2msY}
%\end{align}
In order to compute the engtanglement entropy for a strip of width $\ell>0$, we need to compute minimal surfaces that intersect the point $(\mu,y) = (0,\ell)$.  In this low-dimensional case, the surfaces in question are really just geodesics in the geometry \eqref{tsmtwo}.  For the AdS vacuum, one has $f(\mu) =
\csc^2\mu$ with $\mu\in[0,\pi]$, and the calculation  is easy since \eqref{tsmtwo} is the metric on
the Poincar\'{e} upper-half plane in polar coordinates with angular coordinate $\mu$ and
radial coordinate $y$.  In this case, geodesics come in two classes: semicircles centered on
the axis $\mu=0$ and straight lines perpendicular to this axis.  Given a strip of width $\ell$, there is
a family of geodesics consisting of the straight line and semicircles with different radii which
intersect the point $(0, \ell)$.

We are most interested in going beyond the AdS vacuum and examining those functions $f$ that fall into one of the following two families:
those that correspond to a holographic realization of an ICFT, and those that correspond to a
holographic realization of a BCFT.  In both cases, $f$ behaves like $\csc^2\mu$ as $\mu\to 0$ since the spacetime is asymptotically AdS.  In the ICFT case, the geometry flows to another AdS region at some $\mu=\mu_*$ while in
the BCFT case, the geometry becomes singular at some $\mu=\mu_*$.  For such functions $f$,
geodesics in the geometry \eqref{tsmtwo} with initial data imposed near $\mu=0$ behave like
geodesics on the Poincar\'{e} upper-half plane, but they are deformed away from these vacuum solutions as flow move into the
bulk.  Interestingly, for $d=2$ one of the vacuum solutions survives for general $f$.  For any $\ell>0$ and any $f$, the geodesic equation
\eqref{msM} is solved by   $Y(s) = \ell$, the circular solution centered at the origin.  To determine all other relevant geodesics in
the ICFT and BCFT cases, we turn to numerical methods.  We find that with the exception of the circular
solution $Y(s) = \ell$, geodesics in the ICFT and BCFT cases exhibit distinct qualitative
behaviors in the bulk.

\begin{figure}
  \begin{minipage}[c]{0.48\textwidth}
    \includegraphics[scale=0.8]{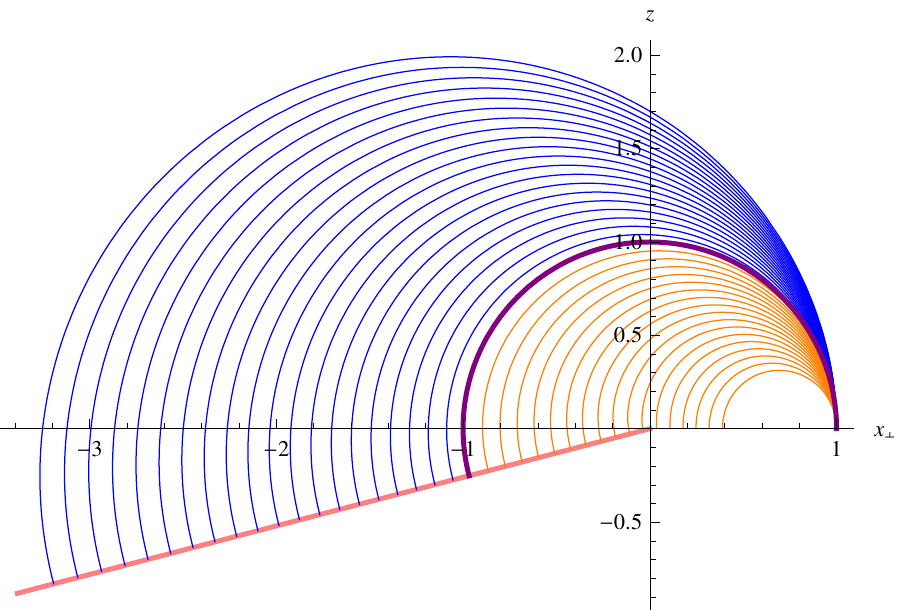}
    \caption{Geodesics for $d=2$ holographic ICFT geometries with $\Delta = 1.212$, $\lambda_4=-4.8$, $\beta_-=0.6$, and $\ell=1$.  The pink radial line indicates the $\mu=\mu_*$ ray where the geometry is asymptotically $\mathrm{AdS}_3$.}
    \label{ICFT2}
  \end{minipage}\hfill
  \begin{minipage}[c]{0.48\textwidth}
    \includegraphics[scale=0.83]{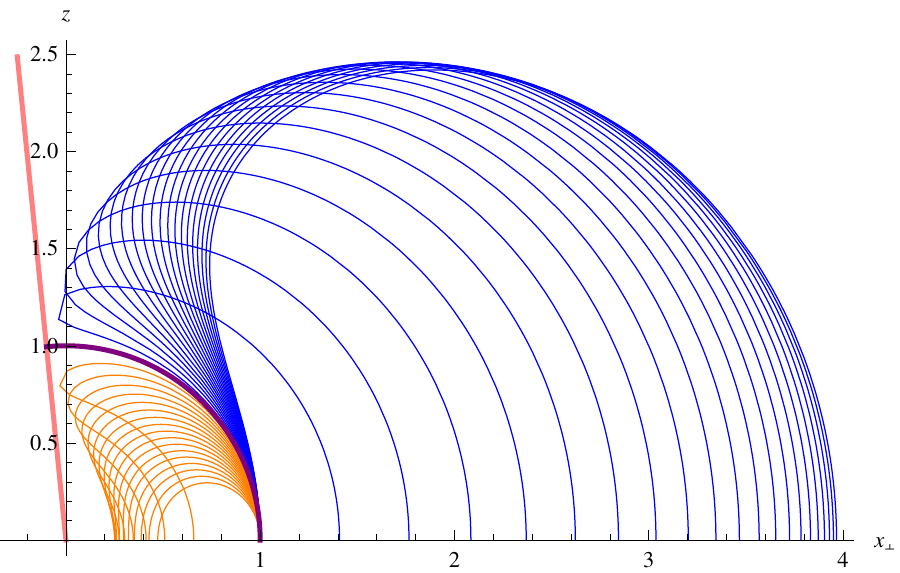}
    \caption{Geodesics for $d=2$ holographic BCFT geometries for $\Delta = 1.212$, $\lambda_4=-4.8$, $\beta_-=1.4$, and $\ell=1$.  The red line indicates the $\mu=\mu_*$ ray where the geometry develops a curvature singularity.}
    \label{BCFT2}
  \end{minipage}
  \vspace{-15pt}
\end{figure}

\subsubsection{ICFT geodesics in $d=2$}

For a given $\ell>0$, there is an infinite family of geodesics intersecting the point $(\mu,y) = (0,\ell)$ on the boundary.  In the parameterization $Y=Y(\mu)$, members of this family are distinguished by the value of the parameter $ \hat   y$ in the asymptotic expansion \eqref{msae}.  In figure \ref{ICFT2}, curves colored orange have negative values of $\hat   y$, while curves colored blue have positive values of $ \hat   y$.  The purple curve has $y=0$ and is the semicircular solution $Y(\mu) = \ell$.  Those curves with a larger value of $y$ intersect the $z=0$ axis at lower values of $x_\perp$.  Solutions that flow all the way to the second AdS region at $\mu=\mu_*$ correspond to entangling surfaces that stretch across the interface.  Those orange solutions that flow back to the first asymptotic region at $\mu=0$ correspond to entangling surfaces that remain on one side of the interface.

%\begin{wrapfigure}[14]{R}{7cm}
%  \includegraphics[scale=0.5]{ICFTPlotCorrect3.pdf}
%  \caption{Geodesics for $d=2$ holographic ICFT geometries.  The red lines indicate the %$\mu=0$ and $\mu=\mu_*$ rays where the geometry is asymptotically $\mathrm{AdS}_3$.}
%  \label{MinSurPlot7}
%\end{wrapfigure}

\subsubsection{BCFT geodesics in $d=2$}

As in the ICFT case, for a given $\ell>0$, there is an infinite family of geodesics intersecting the point $(\mu, y)= (0,\ell)$, and they are differentiated by the parameter $\hat   y$ in \eqref{msae}.  Unlike in the ICFT case, the geometry exhibits a curvature singularity at $\mu=\mu_*$, and this affects the geodesics.  In particular, there is exactly one geodesic that reaches the singularity: the circular solution $Y(\mu) = \ell$.  Every other geodesic is repelled by the singularity, turns around, returns to the asymptotic AdS region $\mu=0$, and intersects the $\mu=0$ axis. This behavior can be seen in figure \ref{BCFT2}.  As in the ICFT plot, orange curves have $\hat   y<0$ while blue curves have $\hat   y>0$, and the purple curve is the circular solution.

%\begin{wrapfigure}{R}{8cm}
%  \includegraphics[scale=0.43]{BCFTPlotCorrect2.pdf}
%  \caption{Geodesics for $d=2$ holographic ICFT and BCFT geometries.  The red lines %indicate the $\mu=0$ and $\mu=\mu_*$ rays.  In the BCFT case, the geometry becomes %singular at the $\mu=\mu_*$ ray.}
%  \label{MinSurPlot7}
%\end{wrapfigure}

\begin{figure}
  \centering
    \includegraphics[scale=0.80]{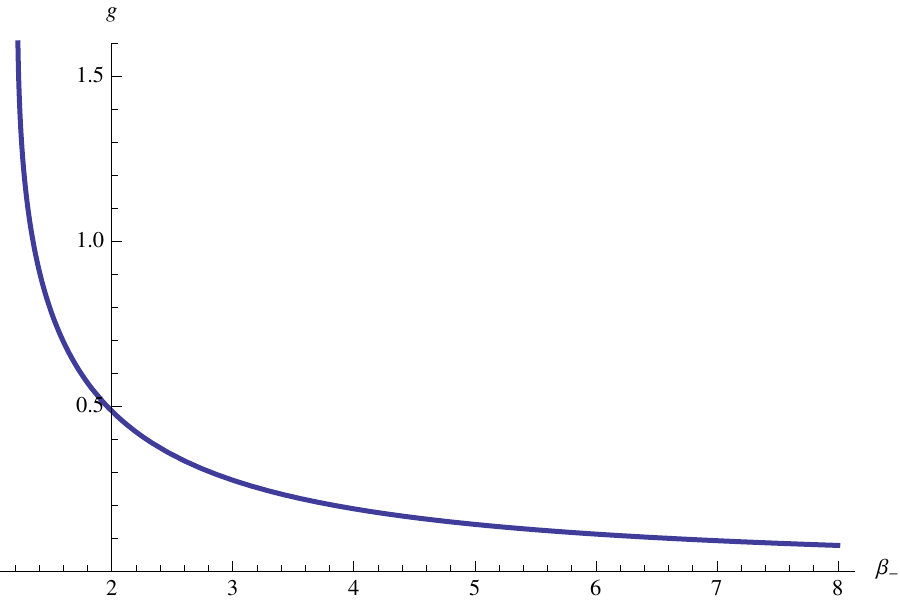}
    \caption{ Boundary entropy ``$g$-factor" $g$ as a function of the source strength $\beta_-$ for a potential with $\Delta=1.212$ and $\lambda_4=-4.8$.}
    \label{boundent}
 \end{figure}

\subsubsection{BCFT holographic entanglement and boundary entropy}

Given a two-dimensional BCFT, it is a well-known result \cite{Calabrese:2004eu}  that the entanglement entropy of the strip geometry is
\begin{align}
    S = \frac{c}{6}\ln\frac{\ell}{\varepsilon} + \ln g \label{2dee}
\end{align}
where $c$ is the central charge of the BCFT, $\ell$ is the width of the strip, $\varepsilon$ is the UV cutoff, $g$ is the so-called $g$-factor introduced in \cite{Affleck:1991tk}, and  $\ln g$ is called the boundary entropy.  Since the symmetric semi-circular geodesic is the only one that reaches the singularity at $\mu=\mu_*$ and therefore the unique one to enclose the boundary of the CFT at the origin, we use it to compute the holographic entanglement entropy, and from this, we can extract the boundary entropy. The area of a minimal surface is computed via \eqref{aread} and \eqref{lagd}.  For the circular solution, we can use the $(\mu, Y(\mu))$ parameterization for the whole curve with $Y(\mu)= \ell$.  The function $f\sim 1/\mu^2$ as $\mu\to 0$ signaling a UV divergence that must be regulated.  In the Poincar\'{e} slicing \eqref{adspoinc}, the UV regulator can be taken as a hard cutoff at some small $z=\varepsilon$.  The coordinate transformation \eqref{adsmap} shows that the appropriate corresponding $\mu$ used to cutoff the area integral is $\mu_\varepsilon=\varepsilon/\ell$ for small $\varepsilon$.  The minimal surface therefore ranges over values of $\mu$ satisfying $\mu_\varepsilon<\mu<\mu_*$, and we obtain the following expression for the holographic entanglement entropy in the strip geometry:
\begin{align}
    S=\frac{1}{4 G_N^{(3)}}\int_{\mu_\varepsilon}^{\mu_*} d\mu\,\sqrt{f(\mu)}.
\end{align}
The series expansion of $f$ about $\mu=0$ shows that after performing the integral, the only divergent piece is that coming from the leading behavior $f\sim 1/\mu^2$ in the AdS region.  In fact, the divergent part is precisely $\log(\ell/\varepsilon)$ as expected from the formula \eqref{2dee}.  Therefore, the boundary entropy can be identified as
\begin{align}\label{gtwod}
    \ln g =\lim_{\varepsilon\to 0}\left[\frac{1}{4 G_N^{(3)}}\left(\int_{\mu_\varepsilon}^{\mu_*}d\mu\,\sqrt{f(\mu)}
    -\ln\frac{\ell}{\varepsilon}\right)\right].
\end{align}
For the  BCFT solution presented in section \ref{twodnumerics} the boundary $g$ factor given in (\ref{gtwod}) can be evaluated numerically.  For the $d=2$ RG flow BCFT found in section  \ref{twodnumerics}, the solution depends on the strength $\beta_-$ of the operator source.
In figure \ref{boundent} we plot the $g$ factor as a function of $\beta_-$  for the numerical example given in section \ref{twodnumerics}.

\subsection{Holographic Entanglement entropy in $d=4$}

The qualitative features of the AdS sliced RG flow solutions in two and four dimensions are very similar.  In this section we solve the minimal surface equations in the GPPZ flow solutions
presented in section \ref{fdourflow}, and we calculate the holographic entanglement entropy for the strip geometry.

Inspection of the minimal surface equation \eqref{asymms} shows that, as the term proportional to $(M')^3$ is non-vanishing in $d=4$, the circular solution $Y(\mu)=\ell$ does not describe a minimal surface.  However in $d=4$, the BCFT background admits a solution which serves as the analog of the $d=2$ circular solution; it is the unique solution that satisfies the desired initial condition $Y(0) = \ell$, and it also reaches the singularity.  This solution can be determined numerically by shooting for the appropriate value of the second undetermined parameter $y$ in the asymptotic expansion \eqref{msaeb} with $d=4$.

\begin{figure}
    %Plots were generated in file named "Josh d=4 GPPZ Numerics 7-18" in the draft folder.
  \begin{minipage}[c]{0.48\textwidth}
    \includegraphics[scale=0.87]{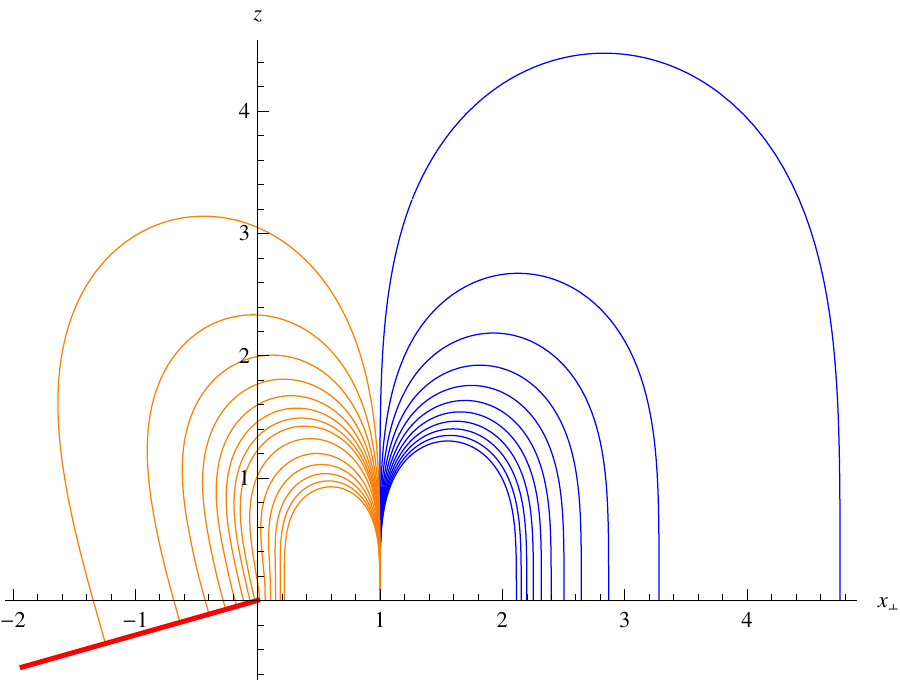}
    \caption{Critical surface profiles for $d=4$ holographic ICFT geometries with $\Delta=3$, $\beta_-=0.3$, and $\ell=1$.  The red line indicates the $\mu=\mu_*$ ray where the geometry is asymptotically $\mathrm{AdS}_5$.}
    \label{ICFT4}
  \end{minipage}\hfill
  \begin{minipage}[c]{0.48\textwidth}
    \includegraphics[scale=0.73]{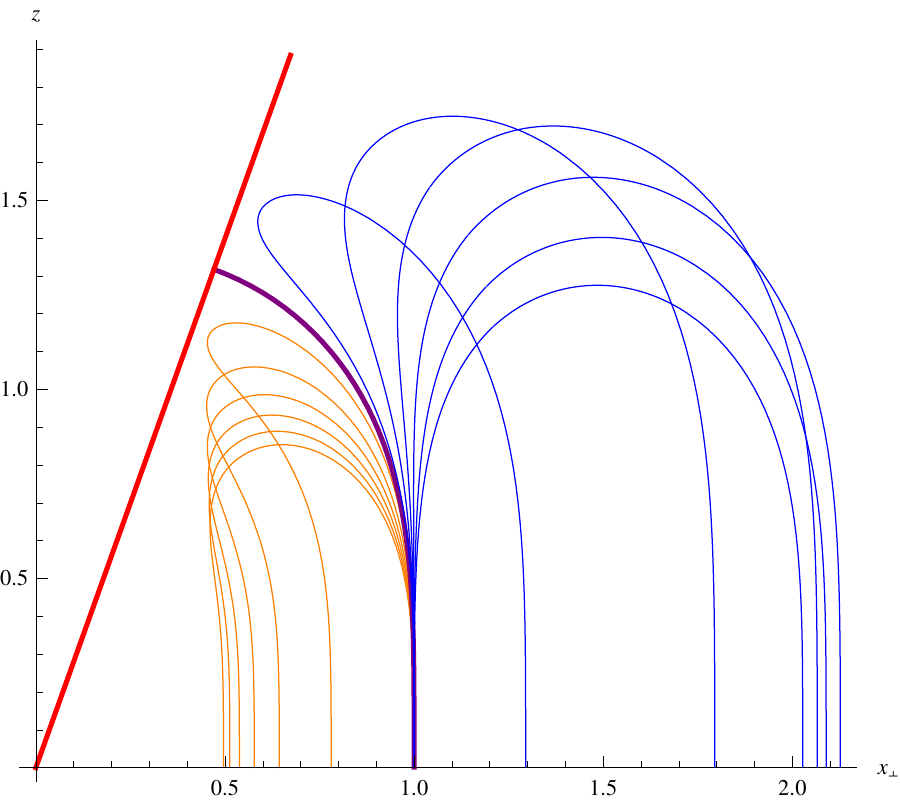}
    \caption{Critical surface profiles for $d=4$ holographic BCFT geometries with $\Delta=3$, $\beta_-=0.6$, and $\ell=1$.  The red line indicates the $\mu=\mu_*$ ray where the geometry develops a curvature singularity.}
    \label{BCFT4}
  \end{minipage}
  \vspace{-15pt}
\end{figure}

\subsubsection{ICFT minimal surfaces in $d=4$}

For each $\ell>0$, there is a family of minimal surfaces satisfying the initial condition $Y(0) = \ell$.  We have plotted this family for the case $\ell=1$ in figure \ref{ICFT4}.  We have identified two subfamilies with colors orange and blue.  The orange curves are solutions with $y$ less than a critical value $y^{(\mathrm {crit})}$.  These solutions either flow back to the asymptotic AdS region at $\mu=0$ with a final value of $Y$ that is less than $\ell$, or they to the asymptotic AdS region at $\mu=\mu_*$.  The blue curves are solutions with $y<y^{(\mathrm {crit})}$. These solutions all flow back to the asymptotic AdS region $\mu=0$ with a final value of $Y$ that is greater than $\ell$.

\subsubsection{BCFT minimal surfaces in $d=4$}

As in the ICFT case, for each $\ell>0$ there is an infinite family of minimal surfaces satisfying the initial condition $Y(0)=\ell$.  We have plotted this family for the case $\ell=1$ in figure \ref{BCFT4}.  Again, we have identified two subfamilies with colors orange and blue which correspond to solutions with $y<y^{(\mathrm {crit})}$ and $y>y^{(\mathrm {crit})}$ respectively.  In addition, we have plotted a purple curve that corresponds to $y = y^{(\mathrm {crit})}$.  This curve is the analog of the $d=2$ circular solution $Y(\mu)=\ell$ in that it is the unique solution reaching the singularity given the initial data $Y(0) = \ell$.

\subsubsection{Holographic entanglement entropy for critical solution}

In this section we will calculate the holographic entanglement entropy for the critical BCFT curve obtained in the previous section.  The entanglement entropy for an RG flow geometry $f(\mu)$ and the critical curve $Y(\mu)$ is given by
\begin{align}
    S=\frac{L^2}{4 G_N^{(5)}}\int_{\mu_\varepsilon}^{\mu_*} d\mu \; {f(\mu) ^{3/ 2} \over Y(\mu) ^3} \sqrt{Y(\mu)^2+ Y'(\mu)^2},
\end{align}
where as in the case $d=2$, $\mu_\varepsilon=\varepsilon/\ell$ for small $\varepsilon$. Due to the singular behavior of $f$ near $\mu=0$, the expression for $S$ is divergent.  Using the expansion around $\mu=0$ given in eq \eqref{dfourexp} and \eqref{msaeb}, one can extract the divergent pieces, and one finds that there is a quadratically divergent and logarithmically divergent contribution with respect to the cutoff $\varepsilon$.  One can define a regular, finite part of the entanglement entropy by subtracting the appropriate divergent terms and then taking $\varepsilon\to 0$;
\begin{align}\label{sregul}
S_{\rm reg} &=  \lim_{\varepsilon\to 0}\left[\frac{L^{2}}{4 G_N^{(5)}}\left(\int_{\mu_\varepsilon}^{\mu_*} d\mu \;   {f(\mu) ^{3/ 2} \over Y(\mu)^3} \sqrt{Y(\mu)^2+ Y'(\mu)^2} -{1\over  2 \varepsilon^2} +{\beta_-^2 \over 3 \ell^2} \log \frac{\ell}{\varepsilon}\right)\right].
\end{align}
 Note that the logarithmically divergent term depends on the source $\beta_-$ of the operator deformation.  We have evaluated the subtracted finite part of the entanglement entropy as a function of $\ell$ and $\beta_-$.  \begin{figure}[h]
  \centering
\includegraphics[scale=0.90]{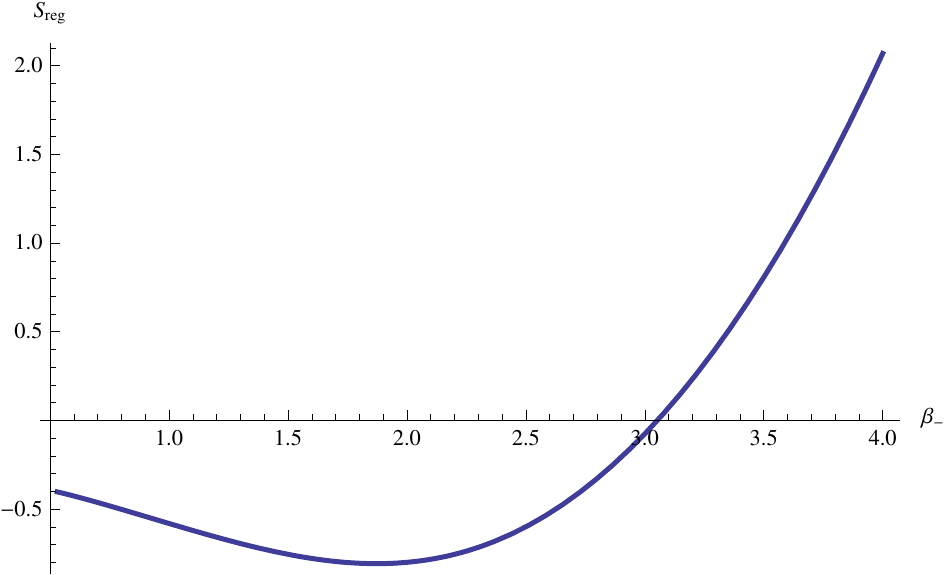}
  \caption{The subtracted finite entanglement entropy as a function as a function of $\beta_-$ for fixed $\ell= 2$.}
\label{a1lplot}
\end{figure}
The numerical results for $S_{\rm reg}$ are well approximated by a  $1/\ell^2$ dependence for any value of $\beta_-$.  This was to be expected since the only dimensionful parameter on which $S_\mathrm{reg}$ can depend is the strip width $\ell$, and the dependence must be $1/\ell^2$ as can be seen from \eqref{sregul}. The dependence on the operator source
is more complicated and reasonably well-approximated by a  quadratic polynomial in $\beta_-$. A representative plot is presented in Figure \ref{a1lplot}.
 It is an open and interesting question whether either the logarithmically divergent
 or finite term are universal and can be interpreted analogously to the $g$-factor in the $d=2$ system. Note that the integration constant $\beta_-$ determines the value
 of $\mu$  where the geometry becomes singular and is therefore equivalent to the
 tension of the cut-off brane in the Takayanagi realization of BCFT. We leave investigations of these questions for future work.

%\begin{align}\label{sregul}
%S_{\rm reg} &=  \frac{L^{2}}{4 G_N^{(5)}}\int_{\mu_\varepsilon}^{\mu_*} d\mu \;  \left( {f(\mu) ^{3/ 2} %\over Y(\mu)^3} \sqrt{Y(\mu)^2+ Y'(\mu)^2} -{1\over  2 \mu_\varepsilon^2 \ell^2} -{a_1^2 \over 3 \ell^2} %\log \mu_\varepsilon\right)
%\end{align}

%%%%%%%%%%%%%%%%%%%%%%%%%%%%%%%%%%%%%%%%%%%
%%%%%%%%%%%%%%%%%%%%%%%%%%%%%%%%%%%%%%%%%%%
\section{Discussion}
\setcounter{equation}{0}
\label{sec5}
%%%%%%%%%%%%%%%%%%%%%%%%%%%%%%%%%%%%%%%%%%%
%%%%%%%%%%%%%%%%%%%%%%%%%%%%%%%%%%%%%%%%%%%

In this paper we have constructed a new holographic description of interface and boundary CFTs utilizing an AdS-slicing ansatz for a holographic RG-flow. In the discussion we compare and contrast this construction with other approaches developed recently in the literature. The construction of Takayanagi et al \cite{Takayanagi:2011zk,Fujita:2011fp} (see \cite{Karch:2000gx} for an earlier, closely related construction) also uses an $\mathrm{AdS}_d$ slicing of $\mathrm{AdS}_{d+1}$ like that given in (\ref{janusmet}). The bulk space is cut off by the presence of a brane with $\mathrm{AdS}_d$ world volume at a fixed value of $\mu$, which is determined by the tension of the brane via matching conditions. In the RG-flow solutions found in the present paper the brane is replaced by the  singularity where $f=0$. For the RG-flow solution, the minimal surface that is used for the calculation of the entanglement entropy is  uniquely determined by the strip width $\ell$ on the boundary. This is to be contrasted with the calculation of the entanglement entropy in \cite{Fujita:2011fp}, where there is a one-parameter family of extremal surfaces where the minimal area solution is used to calculate the entanglement entropy.

The  BCFT  RG-flow solutions develop  curvature singularities, hence  the supergravity approximation, which is only valid for small curvatures, breaks down near the singularity.
  This behavior is similar to what is found in many Poincar\'{e}-sliced RG-flow solutions corresponding to relevant operator deformations such as    the GPPZ flow \cite{Girardello:1999bd}.  The interpretation of the Poincar\'{e} RG-flow is that the theory becomes massive and is in a gapped phase. Nevertheless, calculations of  correlations functions, Wilson loops and entanglement entropy are possible as long as the results are dominated by the region far away from the singularity. We  followed the  same assumption  in the calculation of the entanglement entropy for the AdS-sliced BCFT RG flows presented in this paper.

  In some cases the singularities can be resolved by lifting the solution to higher dimensions (see for example  the discussion of the  Coulomb branch in $\mathcal{N}=4$ SYM given in \cite{Kraus:1998hv,Klebanov:1999tb,Freedman:1999gk}).  Another example of regular BCFT solutions in six-dimensional supergravity corresponding to a backreacted solution of self-dual strings ending on three branes in six-dimensions was found in \cite{Chiodaroli:2012vc}.

As already remarked in \cite{DeWolfe:1999cp}, in contrast to the Poincar\'{e}-sliced RG flows, it is not possible to explicitly integrate  the AdS-sliced RG equations of motion in a first order form based on super potential. Therefore the only solutions we were able to find were numerical. It is an interesting question whether it is possible to find analytic solutions as these would be very useful for, e.g. the holographic calculation of correlation functions in the BCFT.   One approach to construct exact solutions, which has been very fruitful in the past,  is  to solve BPS conditions for the existence of backgrounds which preserve a subset of super symmetries  of the AdS vacuum.  This approach has been very successful in constructing half BPS Janus solutions in type IIB \cite{D'Hoker:2007xy,D'Hoker:2007xz}, M-theory \cite{D'Hoker:2008wc,D'Hoker:2009gg}
 and six dimensional supergravity \cite{Chiodaroli:2011nr,Chiodaroli:2009yw,Chiodaroli:2011fn,Chiodaroli:2012vc}. It would be very interesting to explore these methods to construct  BPS solutions for the the relevant deformations related to the ones in the present paper. If such solutions existed, they would correspond to new interface and boundary CFTs which preserved some superconformal symmetries. Exact solutions would also be important to go beyond the numerical evaluation of  the entanglement entropy and hence   clarify the physical interpretation of the divergent and finite terms in the entanglement entropy.

In 2-dimensional CFT renormalization group flows, ICFTs and BCFTs have also been discussed from a purely field theoretic point of view. For recent examples concerning flows of minimal model CFTs see e.g. \cite{Brunner:2007ur,Fredenhagen:2009tn,Gaiotto:2012wh}. Note that in these examples the relevant perturbations do not have the $x_\perp$ dependence as the ones discussed in the present paper.  A holographic description of $x_\perp$-independent BCFT flow will entail an ansatz that is different from the Janus ansatz used here.  On the other hand it would also be interesting to study the $x_\perp$-dependent relevant perturbations on the field theory side. We leave this for investigation in future work.

\bigskip

\noindent{\Large \bf Acknowledgements}

\bigskip

This  work was supported in part by NSF grant PHY-07-57702. We are grateful to Edgar Shaghoulian for useful comments on a draft of this paper and conversations.

\end{document}